\documentclass[aps,amsmath,amssymb,prc,twocolumn]{revtex4-2}

\usepackage{hyperref}
\usepackage{graphicx}
\usepackage{lineno}
\usepackage{dcolumn}
\usepackage{bm}
\usepackage{epstopdf}
\usepackage{amsmath,amssymb}
\usepackage{multirow}
\usepackage{array} 

\newcommand{\sqn}{\sqrt{s_{\mathrm{NN}}} = 200\mathrm{~GeV}}

\newcommand{\pt}{p_\mathrm{T}}
\newcommand{\pp}{\psi_{2}\lbrace\mathrm{pp}\rbrace}
\newcommand{\idhad}{\pi^{\pm}, K^{\pm}\mathrm{,~and}~p(\bar{p})}
\newcommand{\idhadp}{\pi^{+}, K^{+}\mathrm{,~and}~p}
\newcommand{\vtep}{v_2\lbrace\mathrm{ep}\rbrace}
\newcommand{\vtpp}{v_2\lbrace\mathrm{pp}\rbrace}

\newcommand{\ket}{KE_{\mathrm{T}} = m_{\mathrm{T}}-m_{\mathrm{0}}}

\begin{document}

\title{Elliptic flow of charged hadrons in d+Au collisions at $\sqrt{s_{NN}} =$ 200 GeV using a multi-phase transport model}

\author{J. Tanwar}
 \affiliation{
 Physics Department, Panjab University, Chandigarh, India - 160014
}
\author{I. Aggarwal}
 \affiliation{
 Physics Department, Panjab University, Chandigarh, India - 160014
}
\author{L. Kumar}
 \affiliation{
 Physics Department, Panjab University, Chandigarh, India - 160014
}

\author{V. Bairathi}
\affiliation{
 Instituto de Alta Investigación, Universidad de Tarapacá,\\
 Casilla 7D, Arica, 1000000, Chile
}
\author{S. Kabana}
\affiliation{
 Instituto de Alta Investigación, Universidad de Tarapacá,\\
 Casilla 7D, Arica, 1000000, Chile
}

\date{\today}
\begin{abstract}
This study presents a comprehensive analysis of the elliptic flow coefficient, $v_2$, for charged hadrons at mid-rapidity in d+Au collisions at $\sqrt{s_{\mathrm{NN}}} = 200\mathrm{~GeV}$. Utilizing a multiphase transport (AMPT) model in both default and string melting modes, we examine the dependence of $v_2$ on transverse momentum, collision centrality, and particle type. Furthermore, we present $v_2$ across different collision centralities in d+Au collisions at $\sqrt{s_{\mathrm{NN}}} = 200\mathrm{~GeV}$ within the AMPT model with the string melting mode. Our results indicate that the early-stage partonic phase significantly influences $v_2$, as observed by variations in parton scattering cross-section. We also studied the effect of final-state hadronic interactions on $v_2$ by varying the hadronic cascading time. Comparisons with STAR and PHENIX experimental data show that the AMPT model effectively captures the transverse momentum dependence of $v_2$, underlining the importance of parton scattering mechanisms and the need for careful interpretation of experimental results in asymmetric systems.
\end{abstract}


\maketitle

\section{Introduction}
\label{sec:Intro}
The observation of a hot and dense state of matter, dominated by partonic degrees of freedom, known as Quark-Gluon Plasma (QGP), in high-energy heavy-ion collisions is one of the key discoveries at the Relativistic Heavy Ion Collider (RHIC) and the Large Hadron Collider (LHC)~\cite{exqgp1,exqgp2,exqgp3,exqgp4}. In non-central heavy-ion collisions, the initial overlap area has an anisotropic energy density profile in the transverse plane that creates pressure gradients, leading to momentum anisotropies of particle distribution relative to the symmetric flow planes, a phenomenon referred to as collective flow~\cite{flow1}. The collective flow can be quantified using flow coefficients derived from the Fourier decomposition of the azimuthal angle distribution of produced particles~\cite{flow2,flow3}. Among these, the most commonly studied flow coefficient is the elliptic flow, $v_2$.

Several experimental investigations measuring $v_2$ using various techniques have established the formation of a hydrodynamically expanding QGP medium in heavy-ion collisions~\cite{exflow1,exflow2,exflow3,exflow4,exflow5,exflow6,exflow7}. Recent research on heavy-ion collision experiments has gained interest in asymmetric collision systems. At the RHIC, collisions such as p+Au, d+Au, and $\mathrm{^3He}$+Au have been performed, while at the LHC, p+Pb collisions have been recorded to further explore and understand the formation of QGP medium in these asymmetric collision systems. Early measurements of $v_2$ in p+Pb collisions at the LHC~\cite{exflow8,exflow9,exflow10} and p/d/$\mathrm{^3He}$+Au collisions at RHIC~\cite{exflow11,exflow12,exflow13,exflow14,exflow15} indicate a collective hydrodynamic behavior of the medium produced in these collisions. These findings provide evidence of collectivity in asymmetric collision systems~\cite{exflow16}. A recent measurement of $v_2$ by the STAR collaboration in central p+Au, d+Au, and $\mathrm{^3He}$+Au collisions at $\sqn$ suggest that sub-nucleonic spatial fluctuations, which induce intrinsic deformations in the initial geometry, play a crucial role in the development of the observed azimuthal anisotropies~\cite{exflow17}.

These studies are effectively described by hydrodynamical models in asymmetric heavy-ion collision systems~\cite{exflow15,thflow1}. Calculations using a multiphase transport (AMPT) model that incorporates both hadronic and partonic scattering can also qualitatively describe the $v_2$ observed in asymmetric collision systems~\cite{thflow2,thflow3}. In these simulation models, the elliptic anisotropy arises from interactions among medium constituents due to initial collision geometry and fluctuations in sub-nucleonic structure. Alternative explanations for the observed final-state momentum anisotropy, based on initial momentum correlations, have also been proposed~\cite{thflow4,thflow5}. These findings suggest that hydrodynamic-like flow phenomena, typically associated with symmetric nucleus-nucleus collisions, may manifest in asymmetric collision systems under suitable conditions.

In this study, we report the elliptic flow of charged hadrons at mid-rapidity ($|\eta_{\mathrm{c.m.}}| <$ 0.9) in d+Au collisions at $\sqn$ using the AMPT model. We systematically explore the dependence of $v_2(\pt)$ on parton-parton scattering cross-section, hadronic re-scattering time, and different methods of $v_2$ estimation. Section~\ref{sec:method} discusses analysis methods and briefly describes the AMPT model. Section~\ref{sec:results} presents the findings based on these techniques and compares them to available experimental results focusing on asymmetric-system collectivity. Finally, section~\ref{sec:summary} summarizes and concludes the results. 

\section{Methodology}
\label{sec:method}
This section outlines the model framework and the analysis methods used to calculate the charged hadron elliptic flow in d+Au collisions at $\sqn$.

\subsection{AMPT model}
\label{ssec:model}
AMPT is a multi-phase transport model designed to simulate space-time evolution and dynamics of high-energy nuclear collisions~\cite{ampt1}. It includes both hadronic and partonic phases, including the hadronization process, which converts partons into hardons. The AMPT model is mostly used to study collisions involving proton+proton, proton+nucleus, and nucleus+nucleus systems at center-of-mass energies ranging from a few GeV to 200 GeV at RHIC and up to several TeV at the LHC. A good agreement between experimental data and the AMPT model calculations indicates a strong understanding of nucleus-nucleus collisions~\cite{exflow14,eflow18,eflow19,eflow20,eflow21,ampt2,eflow22,ampt3,ampt4}.

The AMPT model consists of four basic components: (1) initial conditions, (2) parton-parton interactions and re-scattering, (3) conversion from partons to hadrons, and (4) interactions among hadrons and hadronic re-scattering. The AMPT model can function in the default mode (AMPT-DEF) and string melting mode (AMPT-SM). The initial conditions of the collisions are generated using the HIJING model through a Monte Carlo Glauber approach. The HIJING model produces minijet partons via hard processes and excited strings through soft processes. The Zhang Parton Cascade (ZPC) model then simulates the partonic interactions and re-scatterings. In the AMPT-DEF mode, partons are recombined with their parent strings following partonic freeze-out and the excited strings are fragmented into hadrons using the Lund string fragmentation model. In AMPT-SM mode, the strings are first melted into partons, and a quark coalescence model is employed to combine all partons into hadrons. Then, the subsequent hadronic interactions and re-scatterings are described using A Relativistic Transport (ART) model. Therefore, the AMPT model provides a convenient way to explore observables in heavy-ion collisions with and without the partonic phase.

In this work, we simulate 100 million minimum bias d+Au collision events at $\sqn$ using version 2.26t9b of the AMPT model employing both the string melting and default modes. Various input parameters are need to be specified in the model, including parameters $a$ and $b$ in Lund symmetric string fragmentation function, parton cross-section ($\sigma_{qq}$) determined by the QCD coupling constant ($\alpha_{s}$) and the parton screening mass $(\mu$), and hadronic rescattering time ($t_{had}$). The Lund symmetric fragmentation function models hadron production from quark and anti-quark fragmentation, taking into account energy, rapidity, and transverse momentum. A higher value of parameter $a$ results in softer fragmentation with lower mean transverse momentum and higher particle multiplicity. Conversely, a lower value of parameter $b$ indicates higher effective string tension, leading to increased mean transverse momentum for quarks and a higher production of strange quarks relative to non-strange quarks~\cite{ampt4}. We used default values of the input parameters, which reasonably described the multiplicity density, particle $\pt$-spectra, and elliptic flow, for Au+Au collisions at $\sqn$~\cite{ampt5,ampt6}. We also varied these parameters, as listed in Table~\ref{tab1}, to systematically study the dependence of $v_2$ on the parameters $\sigma_{qq}$ and $t_{had}$ within the AMPT-SM model.
\begin{table}[!htbp]
\begin{center}
\caption{Input parameters of the AMPT-SM model.}
\vspace{0.2cm}
\label{tab1}
\renewcommand{\arraystretch}{1.20}
\setlength{\tabcolsep}{8pt}
\begin{tabular}{c c c}
\hline
\hline
\multirow{2}{*}{\bf Parameter}	& \multicolumn{2}{c}{\bf Values} \\[0.2ex]
\cline{2-3}
								& {\bf Default} & {\bf Variations} \\[0.2ex]
\hline
$\bf a$             				& 0.55   	& - \\
$\bf b$ (GeV$^{-2}$)				& 0.15    	& - \\
$\bf \sigma_{qq}$ (mb)			& 3.0    	& 1.5, 6.0 \\
$\bf t_{had}$ (fm/$c$)			& 30      	& 0.6 \\
\hline
\hline
\end{tabular}
\end{center}
\end{table}

\begin{figure}[!htbp]
\begin{center}
\includegraphics[width=0.4\textwidth]{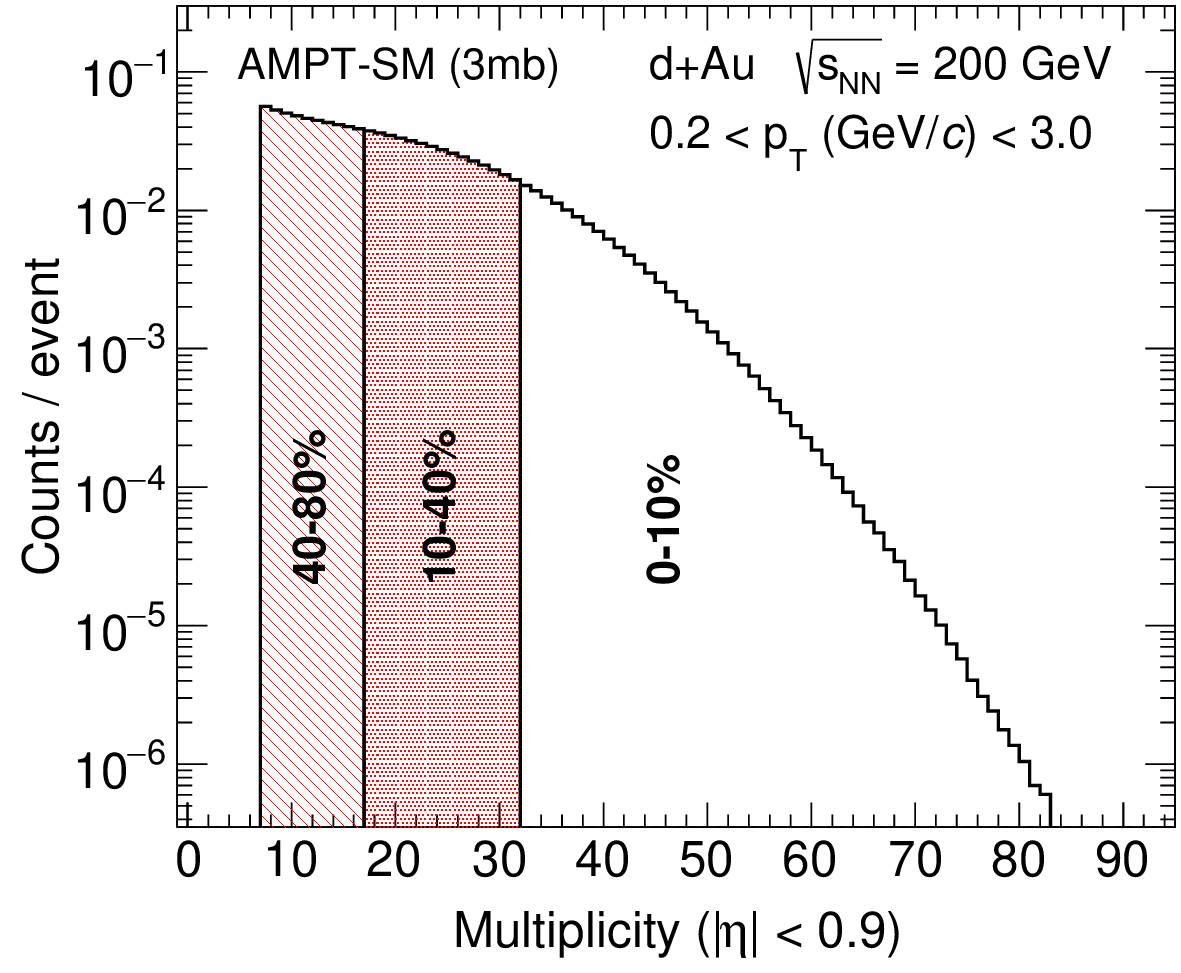}
\end{center}
\caption{(Color online) Charged particle multiplicity distribution at mid-rapidity ($|\eta| < 0.9$) in d+Au collisions at $\sqn$ from the AMPT-SM model.
Different bands show selection of 0-10\%, 10-40\%, and 40-80\% collision centrality.}
\label{fig:refmult}
\end{figure}

The calculation of $v_{2}$ in this study is carried out at various centrality intervals. The charged particle multiplicity distribution is used to determine the centrality of an event. This distribution is obtained within a pseudo-rapidity range of $|\eta|<$ 0.9 and transverse momentum range of 0.2 $< \pt <$ 3.0 GeV/$\it{c}$ similar to that in STAR experiment at RHIC~\cite{exflow17}. Figure~\ref{fig:refmult} represents the multiplicity distribution for d+Au collisions at $\sqn$ from the AMPT-SM model. The events are categorized into three centrality classes: central (0-10\%), mid-central (10-40\%), and peripheral (40-80\%).

\subsection{Flow Analysis Method}
\label{sec:analysis}
The anisotropic flow can be quantified through Fourier decomposition of the azimuthal angle distribution of produced particles with respect to the symmetric collision plane, expressed as follows:
\begin{equation}
\frac{dN}{d\phi} \propto 1 + \sum_{n} 2 v_{n}\cos{\left[(n(\phi -\psi_{n})\right]},
\end{equation}
where $\phi$ denotes the azimuthal angle of the produced particles and $\psi_n$ represents the angle of the n$^{th}$-order symmetric plane of a collision event. The coefficients $v_n$ characterize the strength of anisotropy for each harmonic order n. The second-order coefficient, $v_2$, referred to as elliptic flow, reflects the hydrodynamic response to the initial spatial energy density distribution produced in the early stages of the collision~\cite{flow2,flow3}. 

\begin{figure}[!htbp]
\begin{center}
\includegraphics[width=0.4\textwidth]{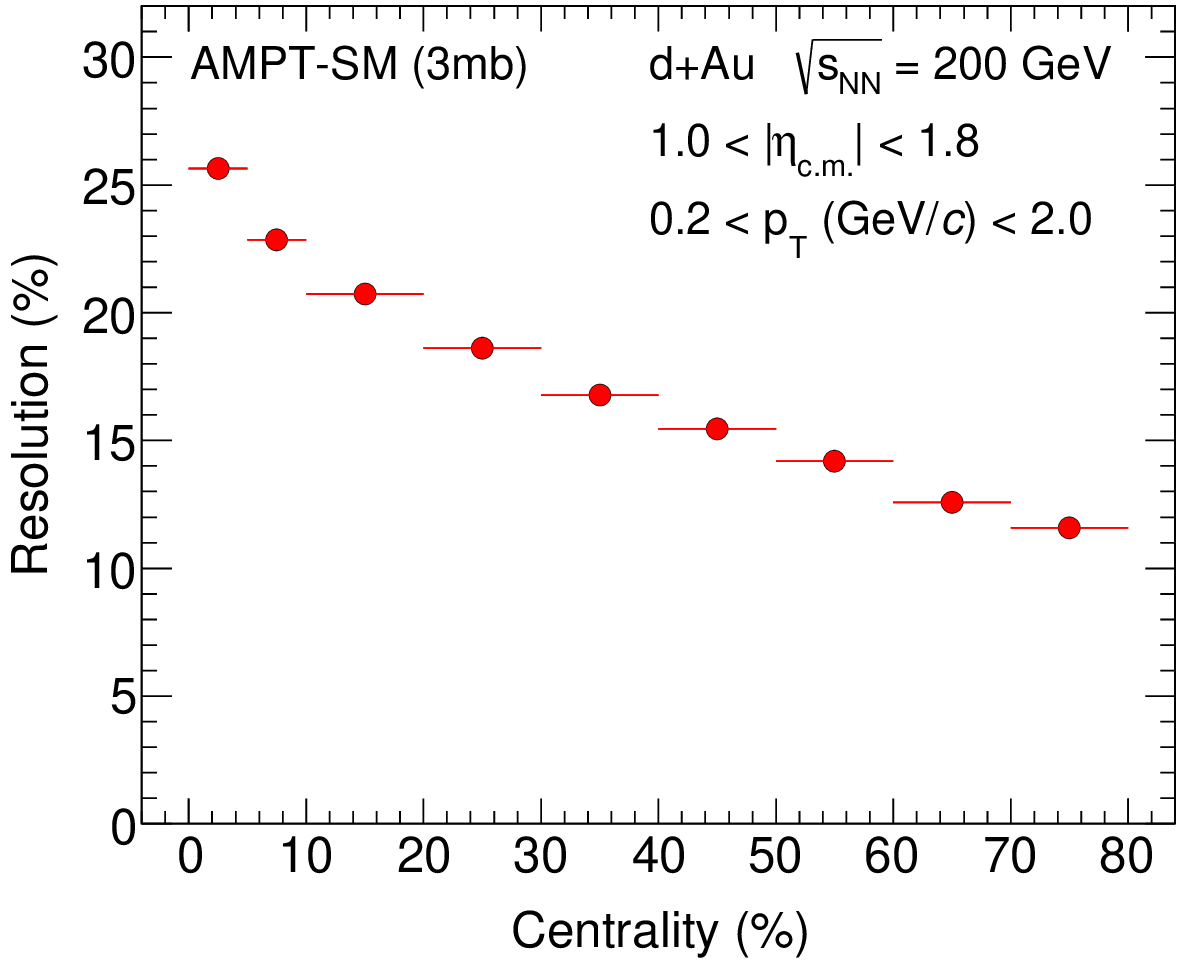}
\end{center}
\caption{(Color online) Event plane angle resolution as a function of centrality using the $\eta$-sub event plane method in d+Au collisions at $\sqn$ from the AMPT-SM model.}
\label{fig:epres}
\end{figure}

In this study, we use $\eta$-sub event plane method to evaluate $v_{2}$ of charged hadrons in d+Au collisions at $\sqn$~\cite{flow3}. We begin with the reconstruction of event plane angle, $\psi_{2}$, in two different sub-events comprised of charged hadrons within a negative pseudorapidity range (-1.8 $< \eta_{\mathrm{c.m.}} <$ -1.0) and a positive pseudorapidity range (1.0 $< \eta_{\mathrm{c.m.}} <$ 1.8) and transverse momentum range of 0.2 $< \pt <$ 2.0 GeV/$\it{c}$, following the approach outlined in Ref.~\cite{exflow17} by the STAR experiment at RHIC. We then obtain $v_{2}$ as a function of $\pt$ for charged hadrons at mid-rapidity ($|\eta_{\mathrm{c.m.}}| <$ 0.9) with respect to the event plane angle derived from opposite pseudorapidity hemispheres as,
\begin{equation}
 v_2^{obs} = \langle\cos[2(\phi-\psi_2)]\rangle,   
\end{equation}
where $\langle \rangle$ denotes average over all particles in all events and $\phi$ is the azimuthal angle of the particle. An $\eta$-gap between hadrons used to calculate $\phi$ and $\psi_{2}$ is applied to reduce non-flow effects due to the short-range correlations. Furthermore, since $\psi_{2}$ is reconstructed from a limited number of produced particles, we correct the observed $v_{2}$ for the $\eta$-sub event plane angle resolution as,
\begin{equation}
v_2 = v_2^{obs} / R_{2}.   
\end{equation}
Here $R_{2} = \sqrt{\langle\cos[2(\psi_2^{\eta+}-\psi_2^{\eta-})]\rangle}, $ is the $\eta$-sub event plane angle resolution. Figure~\ref{fig:epres} shows second-order $\eta$-sub event plane angle resolution as a function centrality in d+Au collisions at $\sqn$ from the AMPT-SM model. The resolution increases from peripheral to central collisions, as it depends on the particle multiplicity.

Due to finite number and fluctuations in the positions of nucleons within the colliding nuclei, the initial collision geometry varies event by event. As a result, the participant plane angle may not necessarily align with the true symmetry plane of the collision. These initial geometric fluctuations can lead to the development of elliptic flow. We also calculated $v_{2}$ relative to the participant plane angle, denoted as $\pp$. The participant plane angle is obtained event-by-event using transverse positions (r, $\phi$) of the participating nucleons in the center of mass frame~\cite{flow4}. 

\section{Results}
\label{sec:results}
\subsection{Transverse momentum dependence of $v_2$}
\label{ssec:v2pt}
\begin{figure*}[!htbp]
\begin{center}
\includegraphics[width=0.7\textwidth]{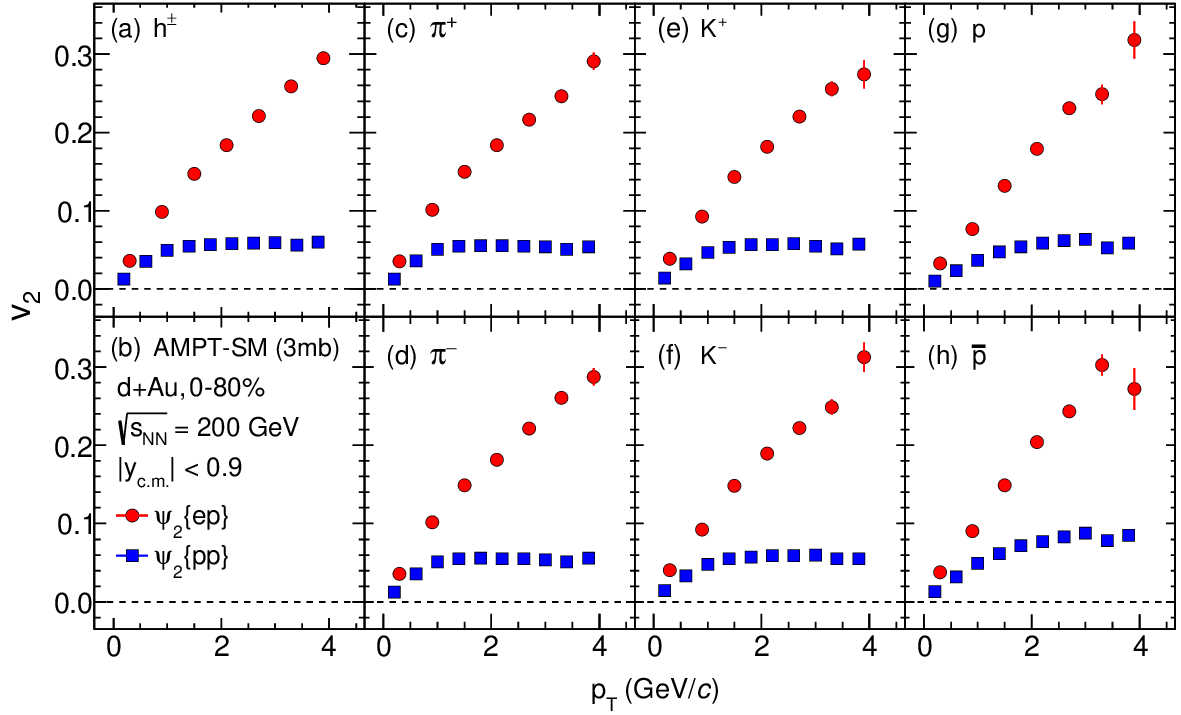}
\end{center}
\caption{(Color online) $v_2$ as a function of $\pt$ for inclusive charged hadrons and identified hadrons at mid-rapidity in 0-80\% central d+Au collisions at $\sqn$ from the AMPT-SM model. The statistical uncertainties are indicated by the error bars.}
\label{fig:3}
\end{figure*}

We evaluated $v_2$ as a function of $\pt$ for inclusive charged hadrons ($h^{\pm}$) and identified hadrons ($\idhad$) at mid-rapidity in 0-80\% central d+Au collisions at $\sqn$. The results are obtained from the AMPT-SM model with a parton-parton cross-section set at 3 mb. Figure~\ref{fig:3} shows $v_2$ calculated with respect to the $\eta$-sub event plane and participant plane angles. The $\vtep$ shows a clear monotonic increase with increasing $\pt$, suggesting a growing contribution of collective motion. In contrast, the $\vtpp$ values gradually increase till $p_{T}\sim$2.0 GeV and then shows a saturation at higher $\pt$. The $\vtep$ exhibits a significantly larger magnitude across all identified particles and inclusive charged hadrons than the $\vtpp$ values.

\begin{figure*}[!htbp]
\begin{center}
\includegraphics[width=0.4\textwidth]{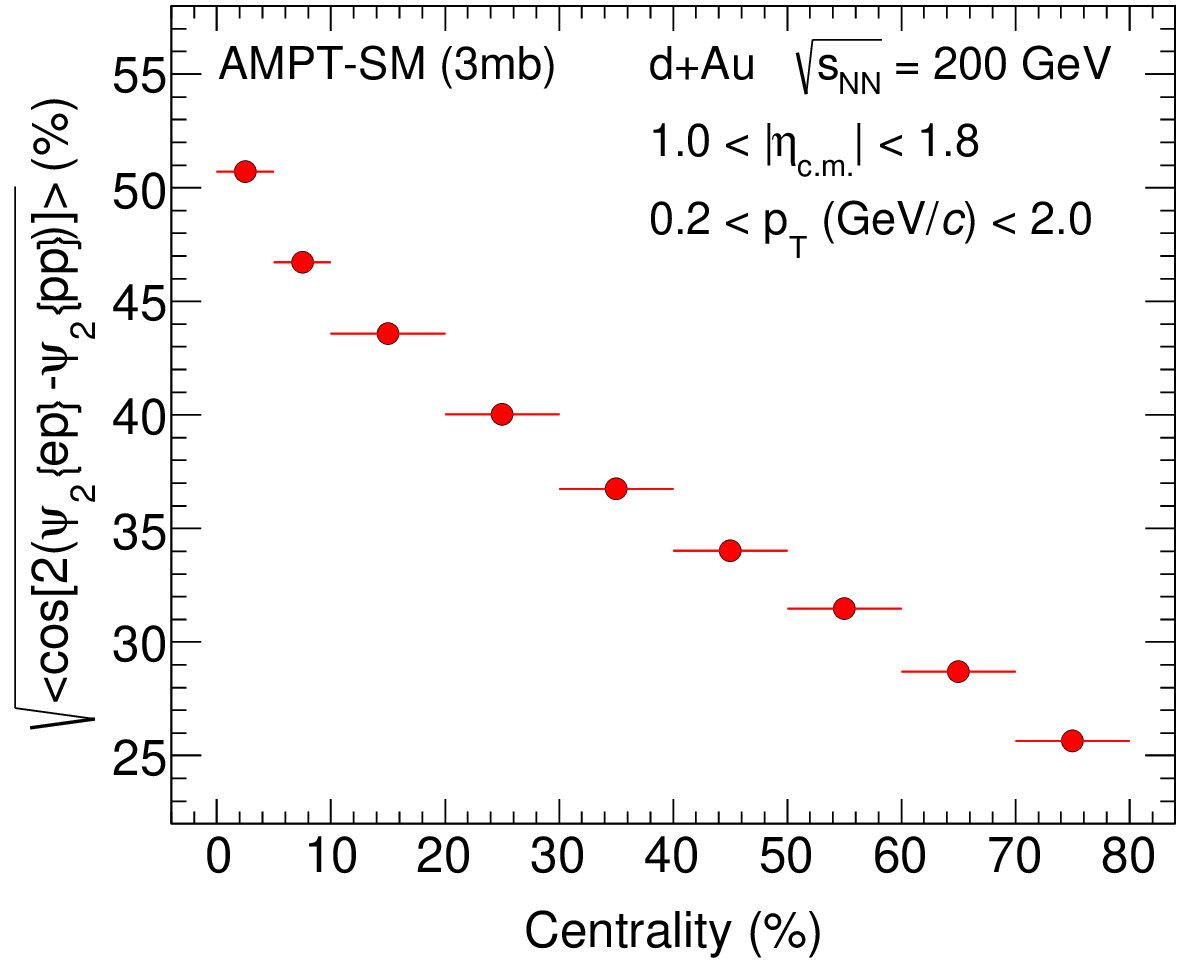}
\end{center}
\caption{(Color online) $\sqrt{\langle\cos[2(\psi_{2}\lbrace{ep\rbrace} - \psi_{2}\lbrace{pp\rbrace}]\rangle}$ (\%) as a function of centrality (\%) in d+Au collisions at $\sqn$ from the AMPT-SM model.}
\label{fig:3b}
\end{figure*}

The observed difference between $\vtep$ and $\vtpp$ in d+Au collisions within the AMPT model can be attributed to the different angles of the event plane and participant plane, as shown in Fig.~\ref{fig:3b}. This difference might results from large event-by-event fluctuations in the final-state charged particles and the number of participating nucleons that are used to calculate the event plane and participant plane angles, respectively.

\subsection{Centrality dependence of $v_2$}
\label{v2e2}
The shape of colliding nuclei and initial collision geometry in heavy-ion collisions influence the expansion and dynamics of the produced medium. The centrality of a collision depends on the initial collision geometry and the shape of colliding nuclei. Studying the centrality dependence of $v_2$ is important, as it can provide insights into the properties of the produced medium, thermalization processes, and hydrodynamic behavior~\cite{exflow2}. Therefore, we studied the centrality dependence of elliptic flow by computing charged and identified hadron $v_2(\pt)$ in three centrality classes (0-10\%, 10-40\%, and 40-80\%) for d+Au collisions at $\sqn$ using the AMPT-SM model. 
\begin{figure*}[!htbp]
\begin{center}
\includegraphics[width=0.7\textwidth]{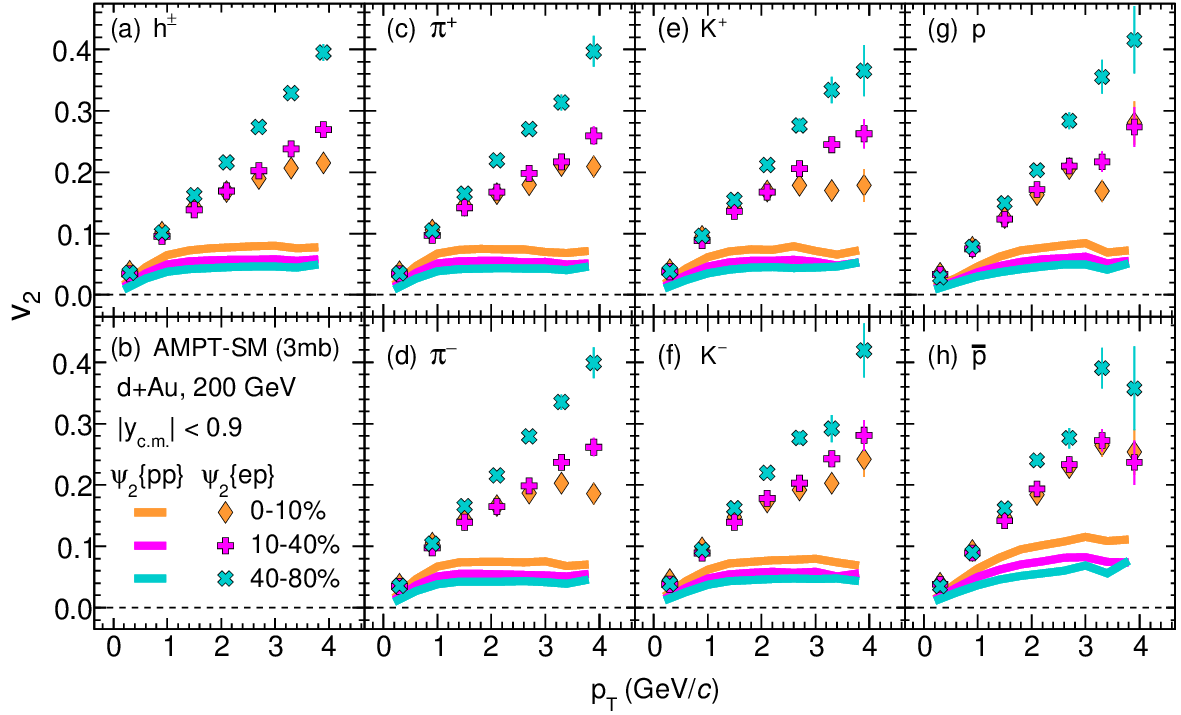}
\end{center}
\caption{(Color online) $v_2$ as function of $\pt$ for inclusive charged and identified hadrons at mid-rapidity d+Au collisions at $\sqn$ from the AMPT-SM model. The statistical uncertainties are indicated by the error bars.}
\label{fig:v2cent}
\end{figure*}

In Fig.~\ref{fig:v2cent}, we present $v_2$ as a function of $\pt$ with respect to $\eta$-sub event plane and participant plane angle. We observed a weak or negligible centrality dependence of $\vtep$ for $\pt <$ 2.0 GeV/$\it{c}$. However, a centrality dependence is observed at $\pt >$ 2.0 GeV/$\it{c}$ in d+Au collisions at $\sqn$, in which the $\vtep$ increases from central to peripheral collisions similar to that observed in symmetric heavy-ion collisions~\cite{exflow2,exflow18,exflow19,thflow6}. The $\vtpp$ exhibit a centrality dependence across all $\pt$ ranges in d+Au collisions at $\sqn$ but the centrality trend is opposite as expected in symmetric heavy-ion collisions. The different trend observed between $\vtep$ and $\vtpp$ might be attributed to the different angles of the event plane and participant plane, which could be related to the large event-by-event fluctuations in the final-state charged particles and the number of participating nucleons that are used to calculate the event plane and participant plane angles, respectively.

\subsection{Particle type dependence of $v_2$}
\label{ncq}
The elliptic flow of identified hadrons in Au+Au collisions at $\sqn$ has been extensively studied at RHIC~\cite{exqgp3,exqgp4}. Large values and centrality dependence of $v_2$, consistent with hydrodynamic model calculations, have been reported~\cite{hydro1,hydro2,hydro3}. The study of identified particle $v_2$ provides valuable insight into the mechanism of particle production in high-energy nuclear collisions. Particle mass ordering of the $v_2$ at low $\pt$ ($<$ 2.0 GeV/$\it{c}$) and baryon-meson separation at intermediate $\pt$ (2-5 GeV/$\it{c}$) suggests a collective expansion of the medium and hadronization through quark coalescence, respectively~\cite{exflow19}. 

The observation of baryon-meson splitting in $v_2$ in heavy-ion collisions suggests that the medium formed is dominated by partonic interactions~\cite{exflow20,exflow21}. The identified hadron $v_2$ when plotted with respect to transverse kinetic energy ($\ket$) scaled by the number of constituent quarks ($n_q$) follows an universal curve. Here, $m_{\mathrm{T}}$ and $m_{\mathrm{0}}$ represent the transverse mass and rest mass of the particles, respectively. The number of constituent quark (NCQ) scaling of elliptic flow reflects the collective behavior of de-confined partons in the early stages of heavy-ion collisions, and it is typically interpreted within the framework of parton recombination or coalescence models~\cite{tncq1,tncq2,tncq3}. In previous studies at RHIC, the scaling of $v_2$ with the number of constituent quarks has been observed in symmetric heavy-ion collisions~\cite{eflow18,exflow19}. However, the absence of baryon-meson splitting and NCQ scaling of $v_2$ in nuclear collisions suggests that the medium formed is dominated by hadronic interactions~\cite{exflow20,exflow21}.

\begin{figure*}[!htbp]
\begin{center}
\includegraphics[width=0.45\textwidth]{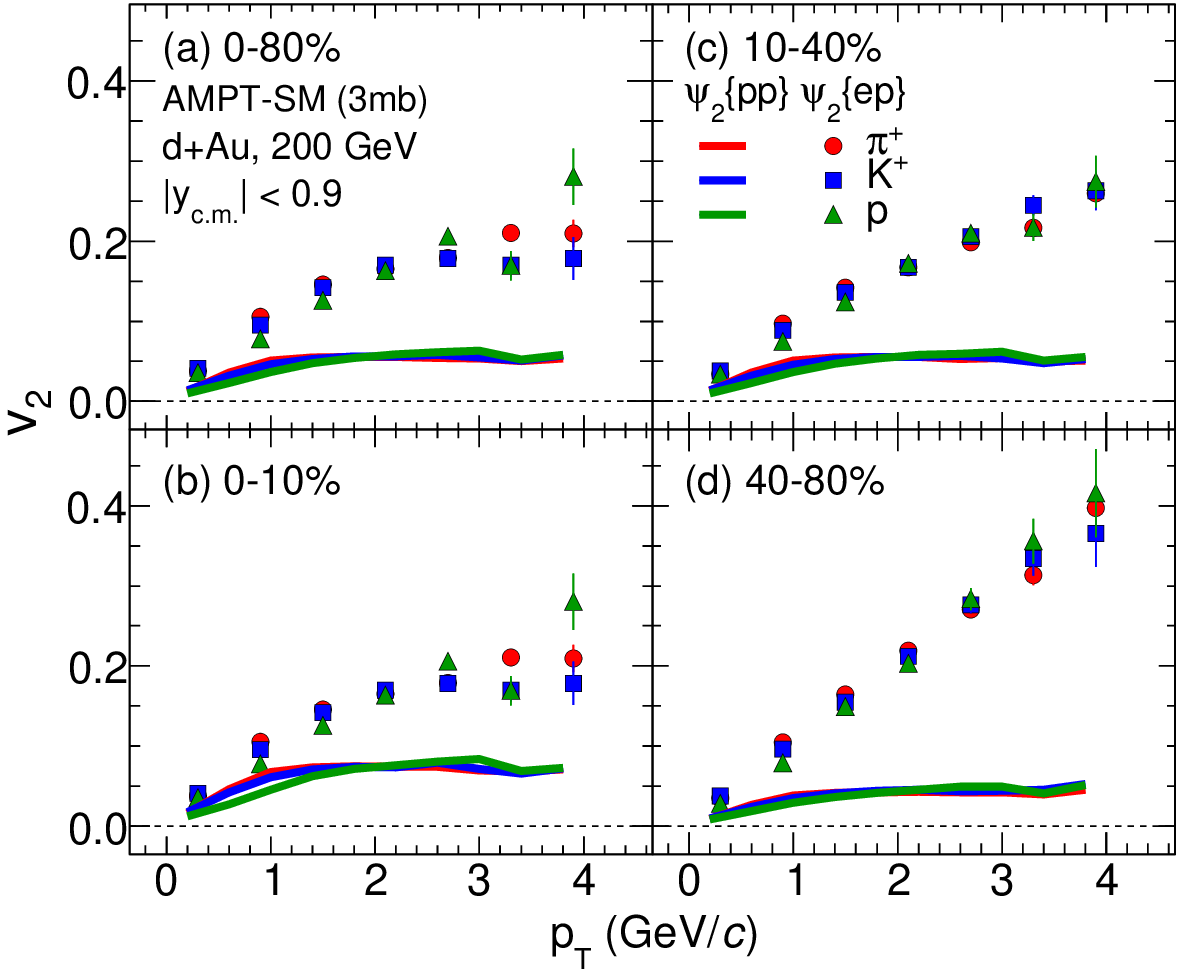}
\includegraphics[width=0.45\textwidth]{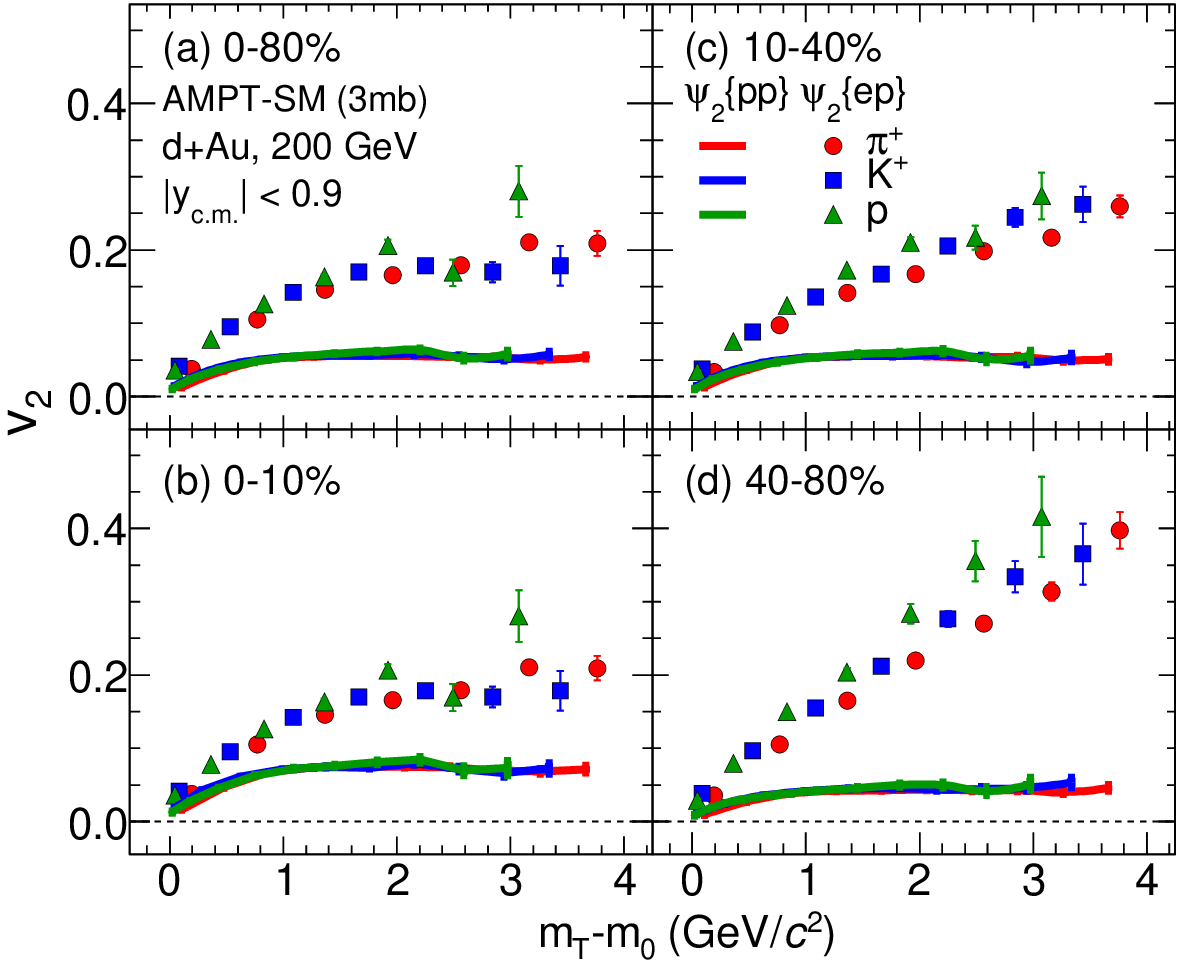}
\end{center}
\caption{(Color online) $v_2$ of $\idhadp$ as a function of $\pt$ (left) and $m_{\mathrm{T}}-m_{0}$ (right) for centrality (a) 0-80\%, (b) 0-10\%, (c) 10-40\%, and (d) 40-80\% in d+Au collisions at $\sqn$ from the AMPT-SM model. The error bars represent statistical uncertainties.}
\label{fig:v2part}
\end{figure*}

We present $v_2(\pt)$ of $\idhadp$ at mid-rapidity in d+Au collisions at $\sqn$ for various centrality classes: 0-80\%, 0-10\%, 10-40\%, and 40-80\% using the AMPT-SM model. The results are shown in Fig.~\ref{fig:v2part} (left) for $v_2$ calculated with respect to the $\eta$-sub event plane and participant plane angle. There is a weak particle mass ordering at low $\pt$, but no evident baryon-meson separation at intermediate $\pt$ observed, for all centrality classes in d+Au collisions at $\sqn$.

\begin{figure}[!htbp]
\begin{center}
\includegraphics[width=0.45\textwidth]{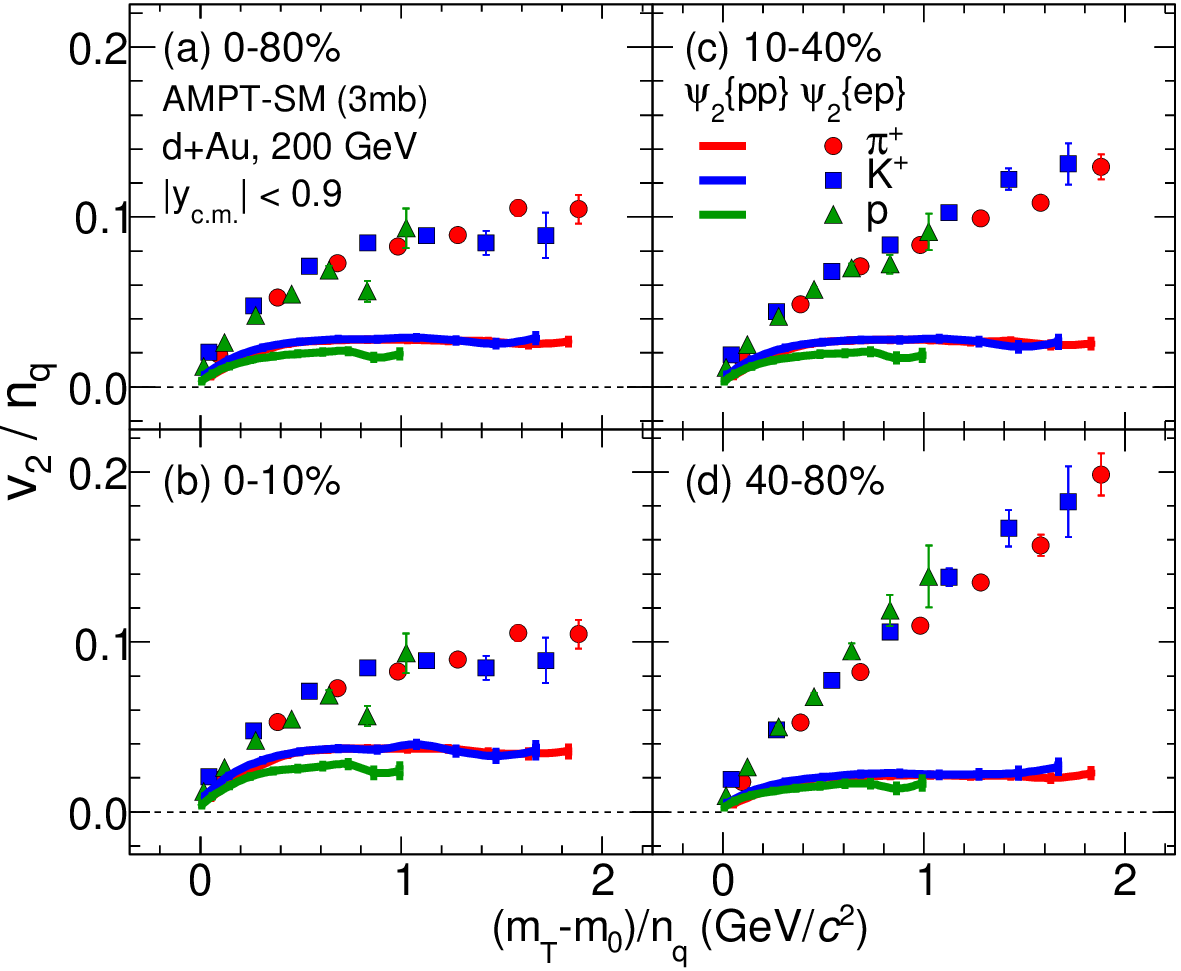}
\end{center}
\caption{(Color online) The NCQ scaled $v_{2}$ as a function of scaled transverse kinetic energy ($m_{T}-m_{0}/n_{q}$) for $\pi^{+}$, $K^{+}$, and $p$ in (a) 0-80\%, (b) 0-10\%, (c) 10-40\%, and (d) 40-80\% in d+Au collisions at $\sqn$ from the AMPT-SM model. The error bars represent statistical uncertainties.}
\label{fig:v2ncq}
\end{figure}

We have studied the identified hadron $v_2$ as a function of $KE_{\mathrm{T}}$ in Fig.~\ref{fig:v2part} (right) to examine the baryon-meson separation in d+Au collisions at $\sqn$. This representation is particularly useful for removing the dependence of particle mass on flow measurements. We observed that $v_2$ plotted against $KE_{\mathrm{T}}$ does not show a clear separation between baryon-meson $v_2$, in contrast to the symmetric heavy-ion collisions~\cite{exflow2}. In Fig.~\ref{fig:v2ncq}, we present $v_2$ of $\pi^{+}$, $K^{+}$, and $p$ scaled by the number of constituent quarks as a function of $KE_{\mathrm{T}}/n_{q}$ for various centrality classes in d+Au collisions at $\sqn$ from the AMPT-SM model. Our analysis shows that we do not observe an adequate NCQ scaling of identified hadron $v_2$ with respect to the event plane and participant plane angle in d+Au collisions at this energy, which is expected as we do not observed the baryon-meson separation of $v_2$. The above results suggest that the system formed in d+Au collision at $\sqn$ might have fewer partonic interactions within the AMPT model framework.

\subsection{Partonic and hadronic scattering dependence of $v_2$}
\label{sigma_qq}
The AMPT model, which incorporates both initial state partonic scatterings and final state hadronic cascading, has been shown to describe collective flow coefficients measured in heavy-ion collisions at RHIC and the LHC. Therefore, the input parameters for parton-parton scattering cross-section and hadronic cascading time in the AMPT model are crucial to understand how hadronization through string fragmentation and quark coalescence mechanism and the subsequent hadronic cascade impact the $v_2$ in asymmetric systems.

\begin{figure}[!htbp]
\begin{center}
\includegraphics[width=0.4\textwidth]{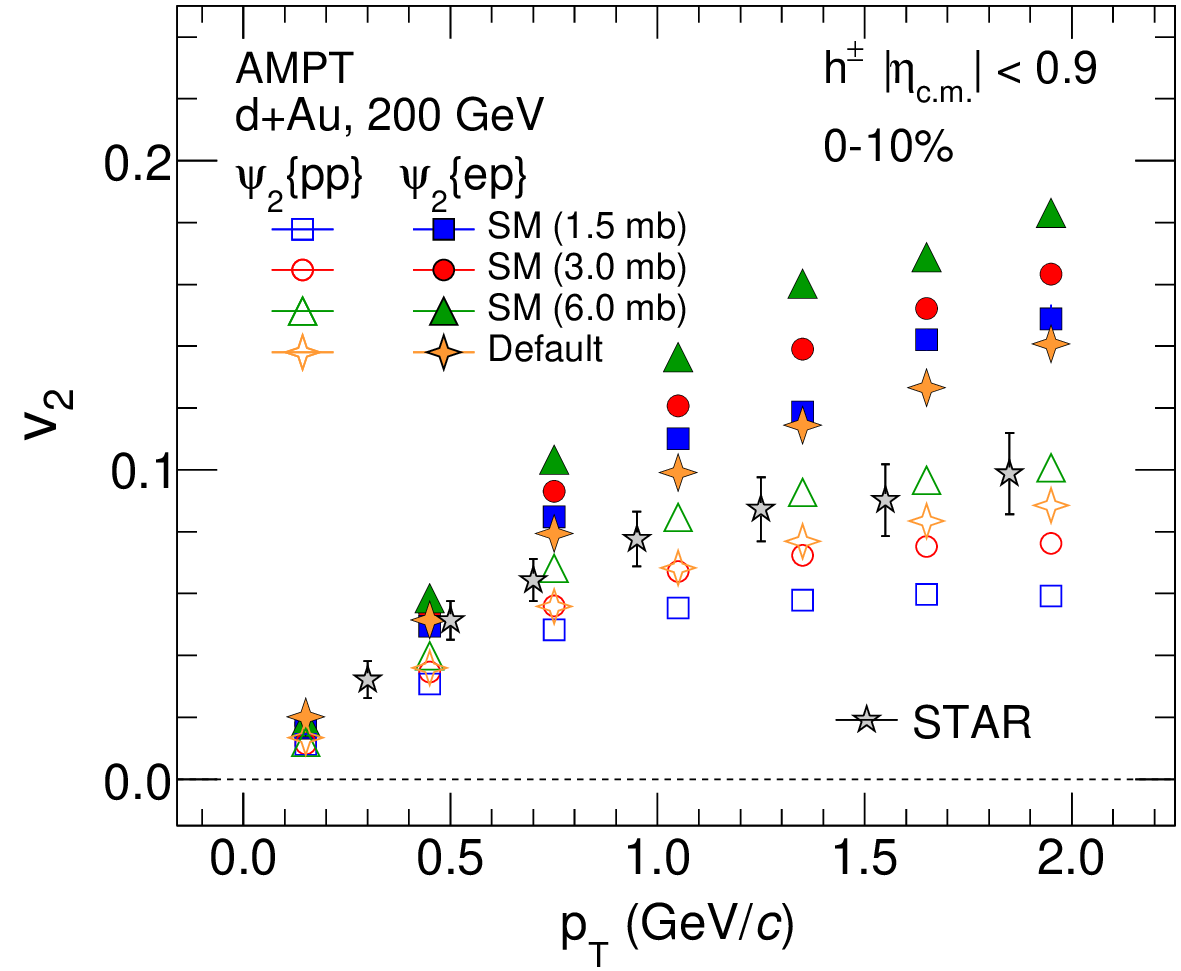}
\includegraphics[width=0.4\textwidth]{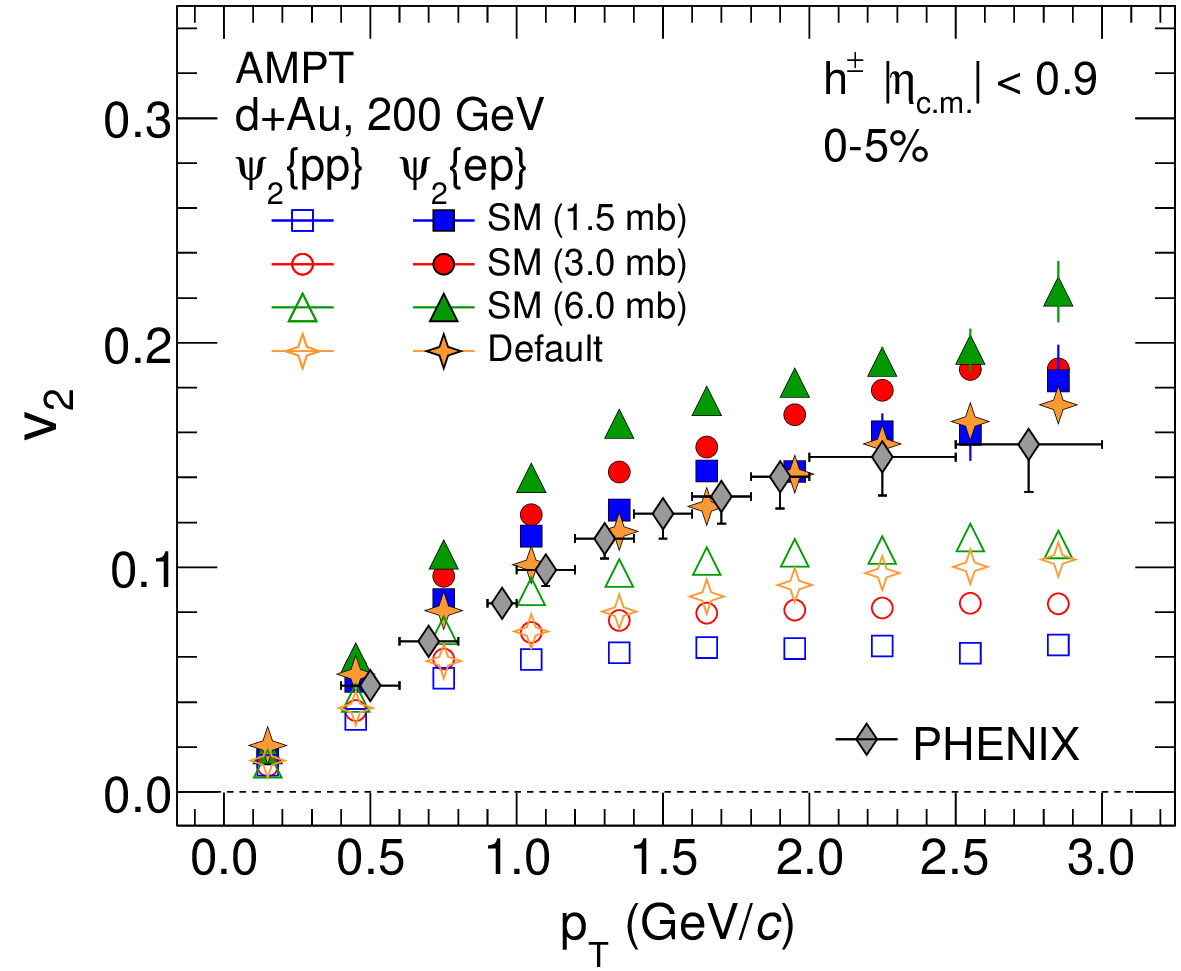}
\end{center}
\caption{(Color online) $v_{2}(\pt)$ of charged hadrons at mid-rapidity in 0-10\% (top) and 0-5\% (bottom) central d+Au collisions at $\sqn$ from the AMPT model. The results of charged hadrons $v_2$ from the STAR and PHENIX experiment at RHIC in same collision system and energy are also shown. Statistical uncertainties are indicated by the error bars.}
\label{fig:sigmaqq}
\end{figure}

In this study, we report charged hadron $v_2$ in d+Au collisions at $\sqn$ using the AMPT-SM model with varying parton-parton cross sections (1.5 mb, 3.0 mb, and 6.0 mb), along with the AMPT-DEF model. Figure~\ref{fig:sigmaqq} shows charged hadron $v_2$ calculated with respect to the $\eta$-sub event plane and participant plane angles with different values of parton scattering cross-sections. We observed that the magnitude of $v_2$ increases with increasing parton scattering cross-section, reflecting stronger partonic collectivity. A larger parton scattering cross-section results in a stronger collective flow signal. This trend is consistent with previous studies in symmetric systems, such as Au+Au and Pb+Pb collisions.

The results from the AMPT model are compared to experimental measurements at mid-rapidity for 0-10\% central d+Au collisions at $\sqn$ from the STAR experiment at RHIC~\cite{exflow17}. We observe that the $v_2$ calculated with respect to the $\eta$-sub event plane angle ($\psi_{2}\lbrace\mathrm{ep}\rbrace$), for all parton scattering cross-section values, overestimates the STAR experimental data. However, $v_2$ calculated with respect to the $\pp$ from the AMPT-DEF model and parton scattering cross-sections of 6 mb from the AMPT-SM model are consistent with the experimental measurements within the uncertainties. Note that the charged hadron $v_2$ from the STAR experiment is measured with the two particle azimuthal angle correlation method, while we use event plane method in the AMPT model.

In Fig.~\ref{fig:sigmaqq}, we also compared our results from the AMPT model with experimental measurements at mid-rapidity for 0-5\% central d+Au collisions at $\sqn$ from the PHENIX experiment at RHIC~\cite{exflow16}. The $v_2$ calculated with respect to the $\eta$-sub event plane from the AMPT-DEF model and parton scattering cross-sections of 1.5 mb from the AMPT-SM model well describe the experimental measurements within the uncertainties. The PHENIX results are obtained using the $\eta$-sub event plane method which provide a critical benchmark for testing flow sensitivity to initial geometry and non-flow suppression techniques. 

It may be noted that the charged hadron $v_2$ values from the STAR and PHENIX experiments differ by approximately 10\%. We note that this discrepancy arises from the use of different methods for subtracting the non-flow. The variation in $v_2$ due to the non-flow effects highlights the significance of various subtraction techniques, which can result in different $v_2$ in small and asymmetric collision systems.

\begin{figure}[!htbp]
\begin{center}
\includegraphics[width=0.4\textwidth]{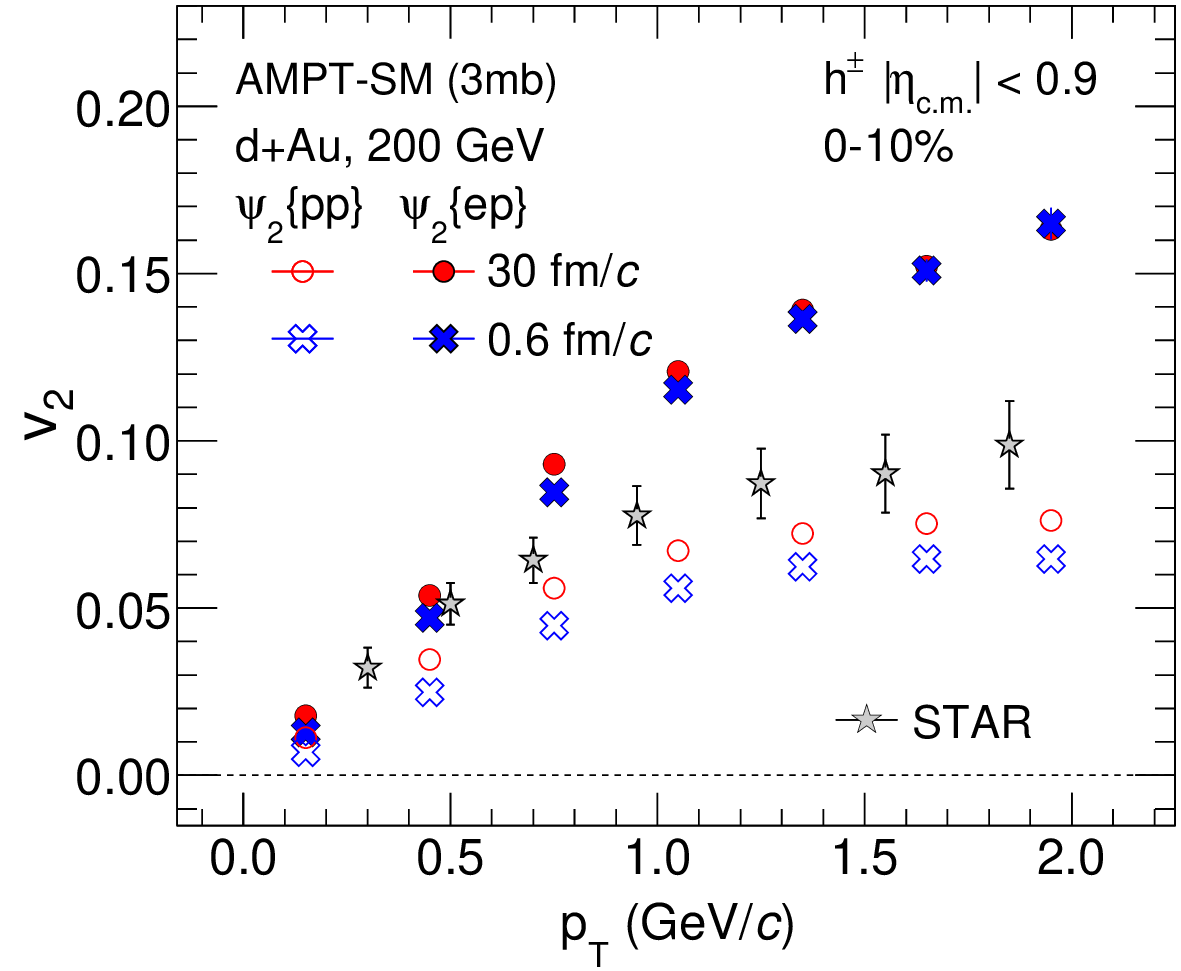}
\end{center}
\caption{(Color online) Charged hadron $v_2(\pt)$ at mid-rapidity in 0-10\% central d+Au collisions at $\sqn$ from the AMPT model with different hadronic cascading time. The results from the STAR experiment at RHIC in d+Au collisions are also shown using gray star markers. Statistical uncertainties are indicated by the error bars.}
\label{fig:thad}
\end{figure}

In Fig.~\ref{fig:thad}, we show a comparison of charged hadron $v_2(\pt)$ at mid-rapidity in 0-10\% central d+Au collisions at $\sqn$ with varying hadron cascading times (0.6 and 30 fm/$c$) and fixed parton scattering cross-section (3 mb) using the AMPT model. This allows us to examine the effect of final-state hadronic interactions and re-scatterings on the calculated $v_2$. The charged hadron $\vtep$ changes slightly with increasing hadron cascading time, indicating that there is a negligible effect on $\vtep$ if we vary hadronic cascading time from 0.6 to 30 $fm/c$. However, the charged hadron $\vtpp$ increases with increasing hadron cascading time implying that it is affected by varying the hadron cascading time in the AMPT model.

The AMPT model predictions are also compared with experimental results from the STAR Collaboration~\cite{exflow17}. The experimental results lie close to $v_2$ calculated with respect to the participant plane angle for a hadron cascading time of 30 fm/$c$ using the AMPT model in d+Au collisions at $\sqn$. 

\section{Summary and Conclusions}
\label{sec:summary}
In this study, we present a comprehensive analysis of the elliptic flow coefficient, $v_2$, for charged hadrons at mid-rapidity in d+Au collisions at $\sqn$. We utilize the AMPT model, both in its default and string melting mode. We discuss the dependence of charged hadrons $v_2$ on particle type, collision centrality, and transverse momentum. To explore the influence of initial conditions, we obtained $v_{2}$ with respect to $\psi_{2}\lbrace\mathrm{pp}\rbrace$ and compared it to the results with respect to $\psi_{2}\lbrace\mathrm{ep}\rbrace$. 

Additionally, we present $v_2(\pt)$ for three different centrality classes (0-10\%, 10-40\%, and 40-80\%) in d+Au collisions at $\sqn$. The study found negligible centrality dependence of $\vtep$ for $\pt$ below 2.0 GeV/$\it{c}$. However, a centrality dependence was observed at $\pt$ above 2.0 GeV/$\it{c}$ in d+Au collisions at $\sqn$, similar to symmetric heavy-ion collisions. The $\vtpp$ showed a centrality dependence across all $\pt$ ranges but exhibited an opposite trend, which indicates large event-by-event fluctuations in final-state charged particles and the number of participating nucleons in d+Au collisions at $\sqn$ compared to symmetric heavy-ion collisions.

We systematically examined the effects of early-stage partonic phase and final-state hadron dynamics on the development of azimuthal anisotropy in asymmetric collision systems by varying the parton scattering cross-section and hadronic rescattering time. The results indicate that increasing parton-parton cross-section in the AMPT-SM model enhances the magnitude of $v_2$. In contrast, the hadronic rescattering time has a minimal effect on the calculated $\vtep$ for inclusive charged hadrons. Our findings highlight the importance of parton scattering mechanisms in shaping the azimuthal anisotropy observed in asymmetric collision systems. 

A comparison of charged hadron $v_2$ in 0-10\% central d+Au collisions with experimental data from the STAR collaboration indicates that the AMPT-SM model, with a partonic cross-section of approximately 6.0 mb, shows better overall agreement in reproducing the transverse momentum dependence of $v_2$ with respect to the participant plane angle. Additionally, a comparison of charged hadron $v_2$ measured with respect to the event plane angle in 0-5\% central d+Au collisions with PHENIX experimental data reveals that the results from both the AMPT-DEF and AMPT-SM models with a small partonic cross-section of 1.5 mb, describe the experimental findings. This discrepancy in $v_2$ results derived from event plane and participant plane angle highlights the significance of model calculations in interpreting the experimental results, especially in asymmetric systems where non-flow effects and event-by-event fluctuations must be carefully considered. 

\section{Acknowledgments}
\label{acknowledgement}
L.K. acknowledges the support of the Research Grant No. SR/MF/PS-02/2021-PU (E-37120) from the Department of Science and Technology, Government of India. S.K. acknowledges the partial financial support of ANID FONDECYT regular No. 1230987 and ANID CCTVal CIA2500027.

\end{document}